\documentclass[prd,aps,preprintnumbers, showpacs, nofootinbib,superscriptaddress, notitlepage]{revtex4}

\usepackage{epsfig}
\usepackage{bm}
\usepackage{amssymb}
\usepackage{amsmath}
\usepackage{color}
\usepackage{subfigure}
\usepackage[colorlinks,
            linkcolor=blue,
            anchorcolor=black,
            citecolor=blue
            ]{hyperref}

\allowdisplaybreaks[4]

\newcommand{\lperp}{\ensuremath{l_\perp}}

\newcommand{\kperp}{\ensuremath{k_\perp}}

\newcommand{\dd}{\ensuremath{\mathrm{d}}}

\begin{document}

\title{Parton shower algorithm with saturation effect}

\author{Yu Shi} 
\affiliation{Key Laboratory of Particle Physics and Particle Irradiation (MOE), Institute of Frontier and Interdisciplinary Science, Shandong University, Qingdao, Shandong 266237, China}

\author{Shu-Yi Wei}
\affiliation{Key Laboratory of Particle Physics and Particle Irradiation (MOE), Institute of Frontier and Interdisciplinary Science, Shandong University, Qingdao, Shandong 266237, China}

\author{Jian Zhou}
\affiliation{Key Laboratory of Particle Physics and Particle Irradiation (MOE), Institute of Frontier and Interdisciplinary Science, Shandong University, Qingdao, Shandong 266237, China}

\begin{abstract}
We extend the previously developed  small $x$ parton shower algorithm  to include the kinematic constraint
effect and $k_t$ resummation effect.  This work enables the Monte Carlo generator to simultaneously resum  large $k_t$ and small $x$ logarithms in the saturation regime for the first time. It is an important step towards simulating processes involving multiple well separated hard scales, such as di-jet production in eA collisions at EIC.
\end{abstract}
\maketitle

\section{ Introduction}
The study of dense gluonic matter at small $x$ inside a large nucleus and nucleon has been and continues to be an important frontier of high-energy nuclear physics. It is also one of the main objectives of the physics program of the future Electron-Ion Collider (EIC)~\cite{Accardi:2012qut,AbdulKhalek:2021gbh}. Tremendous theoretical efforts have been made to search for smoking gun evidence of saturation. To this end, hard scattering processes in eA collisions at EIC are expected to deliver crucial messages about how saturation emerges from strongly interacting gluonic matter. A Monte Carlo event generator that incorporates saturation effects could play an essential role in fully harnessing the potential of future experimental data taken from EIC.

As the core of general purpose Monte Carlo  event generators, parton showers  describe successive radiations from  highly-energetic partons that participate in the hard scattering process.  While most parton branching algorithms~\cite{Ellis:1996mzs,Buckley:2011ms,Hoche:2014rga,Byer:2022bqf} are based on the soft and collinear approximation which effectively resums the Dokshitzer-Gribov-Levin-Altarelli- Parisi (DGLAP)~\cite{Altarelli:1977zs} like logarithm to all orders, only a few parton shower generators~\cite{Marchesini:1990zy, Marchesini:1992jw,Jung:2000hk,Jung:2010si,Andersen:2009nu, Andersen:2009he,CASCADE:2010clj,Andersen:2011hs,Lipatov:2019oxs,vanHameren:2016kkz,Hautmann:2022xuc} have been developed  to describe small $x$ processes by simulating semi-hard emissions which give rise to the logarithm of the type $\ln (1/x)$~\cite{Kuraev:1977fs,Balitsky:1978ic}. Among these generators, the Cascade~\cite{Jung:2000hk,Jung:2010si}  that is built on the  Catani-Ciafaloni-Fiorani-Marchesini (CCFM) evolution equation~\cite{Ciafaloni:1987ur,Catani:1989sg,Catani:1989yc,Marchesini:1994wr} is the most widely used in the phenomenology studies (see for recent examples~\cite{Lipatov:2022doa,Lipatov:2023ypn}). However, none of the aforementioned parton showers takes into account the gluon recombination process occurs in the dense target.

The first attempt to include saturation effect in the parton shower is presented in Ref.~\cite{Shi:2022hee}
where both the  forward and the backward evolution schemes have been presented. The underlying parton branching equation employed in our formulation is the folded Gribov-Levin-Ryskin (GLR) equation~\cite{Gribov:1983ivg}. Although the GLR equation is somewhat outdated compared to modern treatments of small $x$ evolution~\cite{ Balitsky:1995ub,Kovchegov:1999yj,JalilianMarian:1997jx, JalilianMarian:1997gr, Iancu:2000hn, Ferreiro:2001qy}, it is sufficient for simulating events in eA collisions at EIC energy. This is because the gluon density probed at EIC is not high enough for the triple pomeron vertex to dominate the gluon fusion process.  In the previous work~\cite{Shi:2022hee},  we performed a consistent check by comparing the transverse momentum distribution of exchanged gluons reconstructed from the parton shower generator with numerical solutions of the GLR equation.  A full agreement between these two results was reached. The running coupling effect was also implemented in our Monte Carlo simulation.

In the present work, we improve this parton branching algorithm by imposing the  kinematic constraint arising from the requirement that the offshellness of $t$ channel gluon should be dominated by  its transverse momentum squared~\cite{Kwiecinski:1996td, Kwiecinski:1997ee,Deak:2019wms}.  Though it is formally a sub-leading logarithm contribution, the kinematic constraint effect is known to significantly slow down the evolution speed.  It is thus a necessary component of the Monte Carlo generator for any practical phenomenological studies.  Actually, the angular ordering of soft emissions is automatically imposed once the kinematic constraint is applied since the angular ordering constraint is  weaker than the latter~\cite{Kwiecinski:1996td} in the small $x$ limit. The coherent branching effect is thus effectively included in the parton shower. On the other hand, for the case of hard scattering processes involving multiple well-separated hard scales, like di-jet production in eA collisions,  the transverse momentumn dependent (TMD) type large logarithm  $\alpha_s \ln^2 \left(Q^2/k_\perp^2 \right) $ and small $x$ logarithm $\alpha_s \ln \left(1/x\right)$ need to be simultaneously resummed.
Such a joint resummation formalism has been established in a series of publications~\cite{Zhou:2016tfe, Xiao:2017yya,Zhou:2018lfq}. Another main objective of this work is to implement the joint resummation in the Monte Carlo simulation.  

The rest of the paper is organized as follows. In Sec. II, we discuss how to integrate  the kinematic constraint effect into the parton shower algorithm. The formulations of both forward and backward evolution  are presented.  In Sec. III,  the implementation of the joint resummation in the algorithm is discussed. Our starting point is the Sudakov factor derived from a folded version of the  Collins-Soper (CS) and the renormalization group equation.  It is shown that the $k_\perp $ distribution  reconstructed from the  parton shower is identical to the numerical and analytical results obtained from the CS equation and renormalization group equation.  The paper is summarized in Sec. IV.

\section{The kinematic constraint}
In our previous work~\cite{Shi:2022hee},  we have developed a Monte Carlo method to simulate the parton shower at small $x$ based on the GLR evolution equation~\cite{Gribov:1983ivg}. Our formulation only takes into account the summation of the leading logarithm $\ln \left(1/x\right)$ contribution which is known to result in too rapid growth of gluon number density towards small $x$ region. From a phenomenological point of view, it is crucial to go beyond the leading logarithm accuracy and include the various sub-leading logarithm contributions~\cite{Kwiecinski:1996td, Kwiecinski:1997ee,Deak:2019wms, Liu:2022xsc,Avsar:2011ds, Iancu:2015vea, Iancu:2015joa, Lappi:2016fmu,Ducloue:2019ezk, Ducloue:2019jmy,Zheng:2019zul}, among which the
 kinematic constraint effect~\cite{Kwiecinski:1996td, Kwiecinski:1997ee,Deak:2019wms, Liu:2022xsc} is a particularly interesting one. The kinematic constraint is required for the validity of the BFKL/GLR equation at small $x$. The constraint is
needed to ensure that the virtuality of the gluons along the chain is controlled by the transverse momenta. The implementation of the kinematic constraint can significantly slow down the small $x$ evolution and thus lead to a better description of relevant phenomenology. Note that the angular ordering of the gluon emissions is automatically satisfied once the kinematic constraint is imposed in the small $x$ limit. The coherent branching effect is thus effectively achieved following the steps outlined below.

The starting point of the Monte Carlo implementation for such an effect is the folded GLR equation with the kinematic constraint.
Following the arguments made in Refs.~\cite{Kwiecinski:1996td,Deak:2019wms}, the  transverse momentum square of the radiated gluon $l_\perp^2$ must be smaller than $\frac{1-z}{z}k_\perp^2$ where $k_\perp$ and $z$ are transverse momentum and longitudinal momentum fraction carried by the daughter gluon respectively.  
The inclusion of the kinematic constraint leads to a modified GLR equation,
\begin{eqnarray}
\frac{\partial  N(\eta,\kperp)}{\partial \eta }&=&
 \frac{\bar \alpha_s}{\pi}  \int\frac{\dd^{2}\lperp}{\lperp^{2}} 
 N\left ( \eta +\ln\left[ \frac{ k_\perp^2}{ k_\perp^2+ l_\perp^2} \right] , l_\perp+  k_\perp \right ) -
\frac{\bar \alpha_s}{\pi}  \int_0^{k_\perp} \frac{\dd^2 l_{\perp}}{l_\perp^2}  N(\eta,k_{\perp}) - \bar \alpha_{s}  N^{2}( \eta,\kperp),
\end{eqnarray}
with $\bar \alpha_s=\alpha_s N_c/\pi$, $\eta = \ln(x_0/x)$ and $x_0=0.01$. The function $N(\eta,\kperp)$ is related to the normal TMD gluon distribution $G(\eta,k_{\perp})$  through $ N(\eta,\kperp)=\frac{2\alpha_s \pi^3}{ N_c S_\perp}  G(\eta,k_{\perp})$ with $S_\perp$ being the transverse area of nucleon/nucleus. 
Converting the above equation to the folded form of the GLR equation, it reads,
\begin{equation}
\frac{\partial}{\partial \eta } \frac{  N(x,\kperp)}{\Delta_{ns} (\eta ,  k_\perp) } = \frac{ \bar \alpha_s }{\pi} \int _{\Lambda_{\rm cut}} \frac{\dd ^2 l_\perp }{l_\perp ^2} \frac{
 N\left ( \eta +\ln\left[ \frac{ k_\perp^2}{ k_\perp^2+ l_\perp^2} \right] ,\lperp+\kperp \right ) }{\Delta_{ns} (\eta , k_\perp)}.
 \label{eq:bk_kc}
\end{equation}
where $\Delta_{ns} (\eta , k_\perp)$ represents the probability of evolving from $\eta_0$ to $\eta$ without resolvable branching. It is given by,
\begin{eqnarray}
\Delta_{ns} (\eta , k_\perp) = \exp \left \{-\bar \alpha_s \int^\eta_{\eta_0}  d\eta' \left [  \ln\frac{k_\perp^2}{\Lambda_{\rm cut}^2}   +N(\eta',k_{\perp}) \right ]  \right \},
 \end{eqnarray}
 where the infrared cut off $\Lambda_{\rm cut}$ is the  matter of choice about what we classify as a resolvable emission.
Emitted gluons with transverse momentum $l_\perp<\Lambda_{\rm cut} $ are considered as the unresolvable ones. And their contribution has been combined with the virtual correction to cancel the infrared divergence. The resolvable branchings are defined as emissions above this range.  All order contributions from the virtual correction and the unresolvable real emission are resummed into  $\Delta_{ns} (\eta , k_\perp)$ which reduces to the non-Sudakov form factor~\cite{Kwiecinski:1996td}  in the dilute limit by neglecting the saturation term. 
 Eq.~\ref{eq:bk_kc} can be converted into an integral form,
\begin{equation}
N(\eta ,\kperp)=  N(\eta_0 ,\kperp) \Delta_{ns} (\eta ,\kperp)
 +
 \frac{\bar \alpha_s }{\pi }  \int^{\eta} _{\eta_0 }\dd \eta^\prime  
\frac{\Delta_{ns} (\eta ,  \kperp) }{\Delta_{ns} ( \eta ^\prime, \kperp) }
 \int _{\Lambda_{\rm cut}} \frac{\dd ^2 l_{\perp} }{l_\perp ^2} N(\eta^\prime+
 \ln\left[ \frac{ \kperp^2}{ \kperp^2+ l_{\perp}^2} \right], \lperp+  \kperp) .
\end{equation}
It is evident that the kinematic constrained small $x$ equation is no longer a local equation. Namely, the increase of gluon number density at rapidity $\eta$ is driven by the gluon distribution at rapidity $\eta +\ln\left[ \frac{ k_\perp^2}{ k_\perp^2+ l_\perp^2} \right] $ rather than that at the same rapidity $\eta$.  The corresponding weighting factor needs to be modified dramatically for the non-local case as shown below.

\subsection{Forward evolution}
With these derived folded evolution equations, we are now ready to introduce the Monte Carlo algorithm starting with the forward evolution case.  For a given initial condition $N(\eta_i,k_{\perp,i})$, the first quantity to be generated by the algorithm is the value of $\eta_{i+1}$. As it has been done in~\citep{Shi:2022hee}, this task can be achieved  by solving the equation, 
\begin{equation}
\mathcal R
= \exp\left[- \bar \alpha_s \int ^{\eta_{i+1}} _{\eta_i} \dd \eta ^\prime \left(  \ln \frac{ k_{\perp,i }^2}{\Lambda_{\rm cut} ^2}  +  N(\eta ^\prime, k_{\perp,i} )  \right) \right],  \label{etai1}
\end{equation}
where $\mathcal R$ is a random number distributed uniformly in the interval [0,1]. Throughout this paper, we always use $\cal R$ to denote such a random number. $N(\eta ^\prime, k_{\perp,i})$ is pre-generated by numerically solving the GLR equation with the kinematic constraint. 

In contrast to the DGLAP evolution, the unitarity is not preserved during the course of the small $x$ evolution. The number of gluons increases after each step of parton branching. 
The generated cascade thus needs to be re-weighted. For instance, if one neglects the saturation effect and kinematic constraint effect, the number of gluons which vanish due to the virtual correction and the unresolved branching is proportional to  $  \bar \alpha_s \int_{\Lambda_{\rm cut} }^{k_{\perp,i}} \frac{\dd l_\perp^2}{l_\perp^2}$, while the number of gluons produced via the real correction is proportional to  $  \bar \alpha_s \int_{\Lambda_{\rm cut} }^{P_\perp} \frac{\dd^2 l_\perp }{l_\perp^2}$ where $P_\perp$ is the UV cutoff, in the same rapidity interval.  The weighting function is given by the ratio of these two contributions ${\cal W} ( k_{\perp,i}) =\ln(\frac{P_\perp^2}{\Lambda_{\rm cut} ^2}) /\ln(\frac{k_{\perp,i}^2}{\Lambda_{\rm cut} ^2} )
$.

\begin{figure}[htb]
\centering
\includegraphics[width=0.4\textwidth]{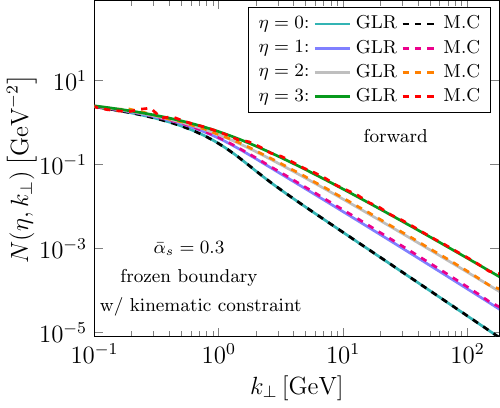}
\includegraphics[width=0.4\textwidth]{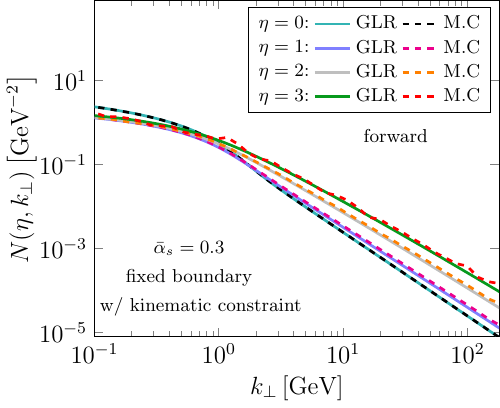}
    \caption{Comparison of the gluon $k_\perp$ distributions reconstructed from the forward evolution approach with the numerical solutions of the kinematic constrained GLR equation at different rapidities. The results obtained with the frozen boundary prescription and the fixed boundary prescription are shown in the left and right plots respectively (Color online). }
    \label{coherent}
\end{figure}

It is quite non-trivial to  work out the correct weighting factor when the kinematic constraint is implemented in the parton branching algorithm.
Let us first discuss the derivation of the weighting factor for the case of the fixed boundary prescription. To work out the correct weighting coefficient,  we first write down the expression for the fraction of gluons at $[\eta_{i+1}, \eta_{i+1}+\delta \eta]$ that come form the branching between  $\eta_{i+1}$ and $\eta_{i} $,
\begin{eqnarray}
&&\delta \eta \frac{\partial}{\partial \eta_{i+1} } \left [  \frac{\bar \alpha_s}{\pi}  \int_{\eta_{i}}^{\eta_{i+1}} \dd \eta' \int_{\Lambda_{\rm cut}} \frac{\dd^2 l_\perp}{l_\perp^2}e^{- \bar \alpha_s \int_{\eta_i}^{\eta'} \dd \eta \left [  \ln\frac{k_{\perp,i}^2}{\Lambda_{\rm cut}^2}   +N(\eta,k_{\perp,i})  \right ] } \theta \left ( \frac{1-z'}{z'} (k_{\perp,i}-l_\perp)^2-l_\perp^2 \right ) \right  ] \nonumber \\&&=\delta \eta  \frac{\bar \alpha_s}{\pi}  \int^{{\rm min}\left [P_\perp,\sqrt {(k_{\perp,i}-l_\perp)^2 \frac{1-z}{z} } \right ]}_{\Lambda_{\rm cut}} \frac{\dd^2 l_\perp}{l_\perp^2}e^{- \bar \alpha_s \int_{\eta_i}^{\eta_{i+1}+\ln \frac{(k_{\perp,i}-l_\perp)^2}{(k_{\perp,i}-l_\perp)^2+l_\perp^2}} \dd \eta \left [  \ln\frac{k_{\perp,i}^2}{\Lambda_{\rm cut}^2}   +N(\eta,k_{\perp,i})  \right ] },
\label{kcw}
\end{eqnarray}
with $z'=x_{i+1} /x'=\exp [\eta'-\eta_{i+1}] $. The kinematic constraint is imposed by the $\theta$-function. Note that the   term originating from the derivative  acting on the integral boundary is equal to 0.  The entire contribution comes from the   derivative  acting on the $\theta$-function.  Meanwhile the fraction of gluons that leave  from the  rapidity interval [$\eta_{i+1}, \eta_{i+1}+\delta \eta$] due to the virtual correction is,
\begin{eqnarray}
\delta \eta \frac{\partial  e^{- \bar \alpha_s \int_{\eta_i}^{\eta_{i+1}} \dd \eta \left [  \ln\frac{k_{\perp,i}^2}{\Lambda_{\rm cut}^2}   +N(\eta,k_{\perp,i})  \right ] }}{\partial \eta_{i+1} }=-\delta \eta \bar \alpha_s\left [\ln\frac{k_{\perp,i}^2}{\Lambda_{\rm cut}^2}   +N(\eta_{i+1},k_{\perp,i}) \right ]e^{- \bar \alpha_s \int_{\eta_i}^{\eta_{i+1}} \dd \eta \left [  \ln\frac{k_{\perp,i}^2}{\Lambda_{\rm cut}^2}   +N(\eta,k_{\perp,i})  \right ] }.
\end{eqnarray}

For the non-local small $x$ evolution, one also needs the input for gluon distribution beyond the small $x$ boundary $x_0=0.01$. There are two common choices for the boundary conditions: i) the  fixed boundary prescription, $N(\eta<0, k_\perp)=0$; ii) the frozen boundary prescription, $N(\eta<0, k_\perp)=N(\eta=0, k_\perp)$. The weighting functions are thus different for different rapidity boundary prescriptions. 

For the fixed boundary prescription, the re-weighting function is given by
\begin{eqnarray}
{\cal W}_{kc,1}(\eta_i,\eta_{i+1};k_{\perp,i}) =\frac{(\eta_{i+1}-\eta_i) \int^{{\rm min}\left [P_\perp,\sqrt { \frac{1-z}{z} (k_{\perp,i} -l_\perp)^2 } \right ] }_{\Lambda_{\rm cut}} \frac{\dd^2 l_\perp}{l_\perp^2} e^{- \bar \alpha_s \int_{\eta_{i+1}}^{\eta_{i+1}+\ln \frac{(k_{\perp,i} -l_\perp)^2}{(k_{\perp,i} -l_\perp)^2+l_\perp^2}} \dd \eta \left [  \ln\frac{k_{\perp,i}^2}{\Lambda_{\rm cut}^2}   +N(\eta,k_{\perp,i})  \right ] }}{(\eta_{i+1}-\eta_i)\ln \frac{k_{\perp,i}^2}{\Lambda_{\rm cut}^2} +
\int^{\eta_{i+1}}_{\eta_i} d\eta
N(\eta, k_{\perp,i})}. 
\end{eqnarray}
Here, the values of $|l_\perp|$ and $\phi_l$ can be generated by solving the following equation
\begin{align}
&
{\cal R} = \frac{1}{\cal C} \frac{\bar \alpha_s}{\pi} \int_{\Lambda_{\rm cut}}^{l_\perp} \frac{\dd^2 l'_\perp}{l_\perp'^2}
\exp\left\{ - \bar \alpha_s \int_{\eta_i}^{\eta_{i+1}+\ln \frac{(k_{\perp,i}-l_\perp')^2}{(k_{\perp,i}-l_\perp')^2+l_\perp'^2}} \dd \eta \left [  \ln\frac{k_{\perp,i}^2}{\Lambda_{\rm cut}^2}   +N(\eta,k_{\perp,i})  \right ] \right\}, \label{eq:dis-lt}
\\
&
{\cal C} = \frac{\bar \alpha_s}{\pi} \int_{\Lambda_{\rm cut}}^{{\rm min} [P_\perp,\sqrt {(k_{\perp,i}-l_\perp')^2 \frac{1-z}{z} } ]} \frac{\dd^2 l_\perp'}{l_\perp'^2}
\exp\left\{ - \bar \alpha_s \int_{\eta_i}^{\eta_{i+1}+\ln \frac{(k_{\perp,i}-l_\perp')^2}{(k_{\perp,i}-l_\perp')^2+l_\perp'^2}} \dd \eta \left [  \ln\frac{k_{\perp,i}^2}{\Lambda_{\rm cut}^2}   +N(\eta,k_{\perp,i})  \right ] \right\},
\end{align}
where ${\cal R}$ again is a random number and ${\cal C}$ is the normalization factor ensuring that the r.h.s. of Eq.~\ref{eq:dis-lt} resides in the region of $[0, 1]$. In the practical Monte Carlo implementation, a veto algorithm is used to be more efficient. Once $|l_\perp|$ and $\phi_l$ are generated, $l$ and $k_{\perp,i+1}$ then can be reconstructed subsequently. We repeat the procedure outlined above until $\eta_{i+1}$ reach  a minimal  cut-off value $\eta_{\rm min}$.  Once the whole cascade is generated, we are able to reconstruct the gluon $k_\perp$ distribution at arbitrary rapidity.

For the frozen boundary case, the weighting factor has to be modified to
\begin{equation}
\mathcal W_{kc,2}(\eta_{i}, \eta_{i+1};k_{\perp,i}, k_{\perp,i+1})
=
\frac{   (\eta_{i+1}-\eta_{i}) \ln \frac{ P_\perp^2}{\Lambda_{\rm cut}^2} }{
   (\eta_{i+1}-\eta_{i})\ln \frac{k_{\perp,i}^2}{\Lambda_{\rm cut}^2}
 +
\int^{\eta_{i+1}}_{\eta_i} d\eta
N(\eta,  k_{\perp,i} )
 }
\frac{ { N(\eta_i + \ln \left[  \frac{   k_{\perp,i+1} ^2	 }{    k_{\perp,i+1}  ^2  +l_\perp^2 } \right] , k_{\perp,i} )}}{  N(\eta_i ,k_{\perp,i}  ) }  ,
\end{equation} 
and the radiated gluon transverse momentum $l_\perp$ is sampled solving the following equation
\begin{equation}
{\cal R} = \frac{1}{\cal C} \frac{\bar \alpha_s}{\pi}  \int^{l_\perp}_{\Lambda_{\rm cut}} \frac{\dd^2 l_\perp'}{l_\perp'^2},
\end{equation}
where the normalization factor for this case is given by ${\cal C} = \frac{\bar \alpha_s}{\pi}  \int^{P_\perp}_{\Lambda_{\rm cut}} \frac{\dd^2 l_\perp'}{l_\perp'^2}$. The $k_\perp$ distribution of the exchanged gluons that directly attaches to the hard part can be reconstructed from the forward evolution algorithm described above. 

Using the recipes described above, we are now ready to generate parton cascade. Following the conventional choice, we use the MV model~\cite{McLerran:1993ni, McLerran:1993ka} result as the initial condition  at rapidity $\eta_0=0$. Since we are interested in simulating events such as di-jet production in eA collisions, it is suitable to utilize  the Weisz$\ddot{a}$ke-Williams (WW) gluon distribution as the initial condition~\cite{Dominguez:2011wm}. It is given by
\begin{align}
 N (\eta_0, k_\perp) = \int \frac{d^2 r_\perp}{2\pi} e^{-i  k_\perp \cdot  r_\perp} \frac{1}{r_\perp^2} \left(1- \exp \bigl[-\frac{1}{4} Q_{s0}^2 r_\perp^2 \ln(e+\frac{1}{\Lambda r_\perp})  \bigr] \right),
\end{align}
with $Q_{s0}^2 = 1$ GeV$^2$ and $\Lambda= 0.24$ GeV.  We explored the behavior of the parton cascade with the both fixed boundary prescription and frozen boundary prescription. From Fig.~\ref{coherent}, one can see that the $k_\perp$ distribution obtained from the forward approach is in perfect agreement with the numerical solutions of the kinematic constrained GLR equation for  both  boundary conditions.

\subsection{Backward evolution}
We now turn to discuss how to implement the kinematic constraint  in the backward evolution which is far more efficient in generating initial state parton shower as compared to the forward approach. The rapidity $\eta_{i+1}$ of gluon participating hard scattering is fixed by external kinematics. $k_{\perp,i+1}$ at the rapidity $\eta_{i+1}$ can be sampled with the distribution $N(\eta_{i+1},k_{\perp,i+1})$, which has to be determined beforehand by numerically solving the evolution equation. The next step is to generate $\eta_{i}$ using a modified non-Sudakov form factor.

The modified non-Sudakov form factor, $\Pi_{ns}$, can be related to the forward non-Sudakov form factor $\Delta_{ns}$ and the gluon distribution $N$ as
\begin{eqnarray}
\Pi_{ns} (\eta_{i+1}, \eta_i; k_{\perp,i+1})=\frac{\Delta_{ns}(\eta_{i+1},k_{\perp,i+1})  N(\eta_i,k_{\perp,i+1})}{\Delta_{ns}(\eta_i,k_{\perp,i+1})  N(\eta_{i+1},k_{\perp,i+1})},
\end{eqnarray}
which looks similar to that derived in our previous work \cite{Shi:2022hee}. However, one has to keep in mind that the gluon distributions appearing in the above formula are obtained by solving the GLR equation with the kinematic constraint.

On the other hand, the non-Sudakov factor can also be expressed as \cite{Shi:2022hee},
\begin{eqnarray}
 \Pi_{ns} (\eta_{i+1}, \eta_i; k_{\perp,i+1})
= \exp \! \! \left [-\frac{ \bar \alpha_s}{\pi}  \int_{\eta_i}^{\eta_{i+1}} \!\ \dd\eta   \int^{P_\perp }_{\Lambda_{\rm cut}} \frac{\dd^2 l_{\perp}}{l_\perp^2}   \frac{  N\left (\eta \!+\!\ln\left[ \frac{ k_{\perp,i+1}^2}{ k_{\perp,i+1}^2+ l_\perp^2} \right]\!, k_{\perp,i+1}+ l_\perp \right )}{ N(\eta,k_{\perp,i+1})} \right ].
\end{eqnarray}

\begin{figure}[htb]
\centering
\includegraphics[width=0.4\textwidth]{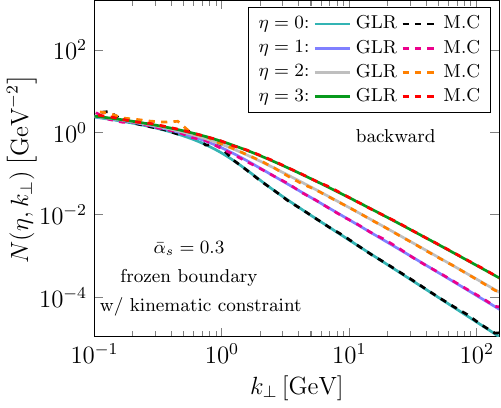}
    \caption{Comparison of the gluon $k_\perp$ distributions obtained   from the backward approach with the numerical solutions of the GLR equation at different rapidities (Color online).}
    \label{coherent_back}
\end{figure}

Both non-Sudakov form factors can be equally well used to generate $\eta_i$ for a given $\eta_{i+1}$ by solving the following equation, 
\begin{equation}
{\cal R} = \Pi_{ns} (\eta_{i+1}, \eta_i; k_{\perp,i+1}).
\end{equation}
The transverse momentum of the radiated gluon $l_\perp$ can be generated according to
\begin{align}
&
{\cal R} = \frac{1}{\cal C} \frac{\bar \alpha_s }{\pi} 
\int_{\Lambda_{\rm cut}}^{l_\perp} \frac{\dd^2 l_\perp'}{l_\perp'^2}   N\left ( \eta_{i+1}+\ln\left[ \frac{ k_{\perp,i+1}^2}{ k_{\perp,i+1}^2+ l_\perp'^2} \right] ,  k_{\perp,i+1}+ l_\perp' \right ),
\\
&
{\cal C} = \frac{\bar \alpha_s }{\pi} 
\int_{\Lambda_{\rm cut}}^{P_\perp} \frac{\dd^2 l_\perp'}{l_\perp'^2}   N\left ( \eta_{i+1}+\ln\left[ \frac{ k_{\perp,i+1}^2}{ k_{\perp,i+1}^2+ l_\perp'^2} \right] ,  k_{\perp,i+1}+ l_\perp' \right ).
\end{align}
Once again, $\cal R$ is a random number, $\cal C$ is the normalization factor and a veto algorithm is employed in our practical implementation to make this sampling procedure more efficient. Similar to the forward evolution case, the generated event has to be re-weighted after each branching in the backward evolution method as well. It is important to notice that the GLR equation with the kinematic constraint is a non-local evolution equation when deriving the weighting factor.  The weighting factor associated with backward evolution is the ratio of the fraction of gluons that appear from branching  at the rapidity $\eta_{i}+ \ln   \frac{   k_{\perp,i+1}^2 }{ k_{\perp,i+1}^2 + l_\perp^2 }$ and the fraction of gluons that vanish at the rapidity $\eta_i$ due to the virtual correction and the fusion process. It reads,
\begin{equation}
\mathcal W_{kc,\text{back}}(\eta_{i+1}, \eta_{i}; k_{\perp,i+1})
=
\frac{
   (\eta_{i+1}-\eta_{i})\ln \frac{k_{\perp,i}^2}{\Lambda_{\rm cut}^2}
 +
\int^{\eta_{i+1}}_{\eta_i} d\eta N(\eta,  k_{\perp,i} )
 }{   (\eta_{i+1}-\eta_{i}) \ln \frac{ P_\perp^2}{\Lambda_{\rm cut}^2} }
\frac{  N(\eta_i ,k_{\perp,i}  ) }{ { N(\eta_i + \ln \left[  \frac{   k_{\perp,i+1} ^2	 }{    k_{\perp,i+1}  ^2  +l_\perp^2 } \right] , k_{\perp,i} )}}  .
\end{equation} 
The procedure outlined above is repeated until $\eta_i$ is smaller than $\eta_0$. The last step of the simulation is to construct four momenta of the radiated gluons. Note that the minus component of the $t$-channel gluon's four momentum can only be reconstructed after the full cascade has been generated. By going from the last $t$-channel gluon (closest to the nucleus), which has the vanishing minus component, forward in the cascade to the hard scattering process, the true minus component of the $t$-channel gluons are constructed. 
 In Fig.~\ref{coherent_back}, we compare gluon $k_\perp$ distribution at different rapidities generated from backward evolution to the numerical solutions of the GLR equation with the kinematic constraint.   
  The perfect match between  gluon $k_\perp$ distributions obtained from the backward approach and by numerically solving the kinematic constrained GLR has been found.

\section{ $k_t$ resummation in the small $x$ limit } 

Our ultimate goal is to build a parton shower generator for simulating events in eA collisions at EIC. The hard scattering processes occurring in eA collisions often involve multiple scales. For instance, loosely speaking, there are three well separated scales in the back-to-back di-jet production: the center mass of energy $\sqrt{s}$, the invariant mass of the di-jet $Q$, and the total transverse momentum of the di-jet system $k_\perp$. To improve the convergence of the pertubative series, the two type large logarithms $\alpha_s \ln \left({s}/{Q^2}\right)$ and $\alpha_s\ln^2 \left({Q^2}/{k_\perp^2}\right)$ arise in the high order calculations of the di-jet production cross section have to be summed to all orders.  The summation of the logarithm contribution $\alpha_s \ln \left({s}/{Q^2}\right)$ is achieved by solving the small $x$ evolution equation, while the  logarithm contribution $\alpha_s\ln^2 \left({Q^2}/{k_\perp^2}\right)$ can be resummed by means of the CS equation. A unified framework that allows us to resum both large logarithms simultaneously in a consistent way have been developed in a sequence of papers~\cite{Zhou:2016tfe, Xiao:2017yya,Zhou:2018lfq}. The evolved
small $x$ gluon TMD  can be expressed as the convolution of the Sudakov form
factor and the renormalized dipole amplitudes. It has been stressed in Refs.~\cite{Xiao:2017yya,Zhou:2018lfq} that at small $x$, gluon TMDs only can be matched onto dipole scattering amplitudes rather than the normal gluon PDFs in the collinear factorization. We notice that such a joint resummation formalism has been studied in the various different context~\cite{Mueller:2012uf,Mueller:2013wwa,Zheng:2014vka,Stasto:2018rci,Marquet:2019ltn,vanHameren:2019ysa,vanHameren:2020rqt,Shi:2021hwx,Zhao:2021kae,Boussarie:2021ybe,Hentschinski:2021lsh,Caucal:2021ent,Taels:2022tza,Caucal:2022ulg,Goda:2022wsc,Benic:2022ixp,Al-Mashad:2022zbq,Goda:2023ryh,Caucal:2023nci,Ganguli:2023joy,Goda:2023jie,vanHameren:2023oiq}.

To simulate hard scattering processes involving multiple scales in a parton shower generator, it is necessary to develop a Monte Carlo branching algorithm to effectively resum both types of logarithms through an iteration procedure. The essential observation that enables the computer implementation of the joint resummation is described as the following. In the backward approach, the evolution starts from the final $t$-channel gluon with the most negative virtual mass-squared, which participates in the hard process. As a parton cascade develops towards the backward direction, the virtual mass of the $t$-channel gluon decreases by radiating soft gluons with the longitudinal momentum fraction $1-z \rightarrow 0$.  This first stage of the evolution is described by the CS equation and the renormalization group equation which resum the double leading $k_t$ logarithm and the single leading  $k_t$  logarithm respectively.  When the virtual mass of the $t$-channel gluon goes down to the scale which is of the order of saturation scale, we should perform the small $x$ evolution. The precise value of this scale should be fixed by fitting the output of the cascade to the experimental data.  During the course of the small $x$ evolution, the virtual mass of the $t$-channel gluon stops monotonously decreasing, whereas its longitudinal momentum fraction increases rapidly until the small $x$ evolution initial boundary is reached.  In this second stage of the evolution, the development of parton cascade is mainly driven by the radiated gluons that carry the large longitudinal momentum fraction $1-z \rightarrow 1$. Therefore, the Monte Carlo algorithm based on the GLR equation should be applied to generate the parton branching at this stage.

To simulate the first stage of the evolution, our primary task is to derive a folded version of the CS equation and the renormalization group equation. To this end, we write down the CS equation in the momentum space,
    \begin{eqnarray}
\frac{\partial N(\mu^2,\zeta^2,\eta,k_\perp) }{\partial \ln \zeta^2}=\frac{\bar \alpha_s }{2 \pi} \int^\zeta_0 \frac{ d^2 l_\perp }{l_\perp^2} \left  [ N(\mu^2,\zeta^2,\eta, k_\perp+l_\perp) -N(\mu^2,\zeta^2,\eta, k_\perp) \right ] \  . 
 \end{eqnarray}
 which can be converted into the conventional expression of the CS equation~\cite{collins2011}  after making the Fourier transform up to the leading logarithm accuracy. Here, $\mu $ is the factorization scale, and $\zeta$ is a scale introduced to regularize the light cone divergence. 
The factorization scale dependence of the gluon TMD in the saturation regime is described by the normal renormalization group equation~\cite{Xiao:2017yya},
   \begin{eqnarray}
\frac{\partial N(\mu^2,\zeta^2,\eta,k_\perp) }{\partial \ln \mu^2}= \bar \alpha_s \left [\beta_0 -\frac{1}{2} \ln \frac{\zeta^2}{\mu^2} \right ]N(\mu^2,\zeta^2,\eta,k_\perp) \  . 
 \end{eqnarray}
with $\beta_0=\frac{11}{12}-\frac{N_f}{6N_c}$ and $N_f=3$ in this work. By choosing the factorization scale $\mu$ to be  $\zeta$, one can combine the CS equation and the renormalization group equation together. The combined evolution equation reads,
   \begin{eqnarray}
\frac{\partial N(Q^2,\eta,k_\perp) }{\partial \ln Q^2}=\frac{\bar \alpha_s }{2 \pi} \int^Q_0 \frac{ d^2 l_\perp }{l_\perp^2}\left [ N(Q^2,\eta, k_\perp+l_\perp)- N(Q^2 ,\eta, k_\perp)\right ] + \bar \alpha_s \beta_0  N(Q^2,\eta, k_\perp), \label{comb}
  \end{eqnarray}
 where $N(Q^2,\eta,k_\perp)\equiv  N(\mu^2=Q^2,\zeta^2=Q^2,\eta,k_\perp)$. Following the standard  procedure, the above evolution equation can be cast into  a folded  equation,
 \begin{eqnarray}
\frac{\partial  }{\partial \ln Q^2} \frac{N(Q^2,\eta,k_\perp)}{\Delta_s(Q^2)}=\frac{\bar \alpha_s }{2 \pi} \int_{\Lambda_{\rm cut}}^{Q} \frac{ d^2 l_\perp }{l_\perp^2}\frac{N(Q^2,\eta, k_\perp+l_\perp) }{\Delta_s(Q^2)}, \label{foldcs}
 \end{eqnarray}
with the Sudakov form factor being given by,
\begin{eqnarray}
{\Delta_s(Q^2)}=\exp \left[ - \int_{ Q_0^2}^{ Q^2}\frac{dt}{t} \frac{\bar \alpha_s (t)}{2} \left ( \ln \frac{t}{\Lambda_{\rm cut}^2}-2\beta_0\right ) \right ].
\end{eqnarray}
The Sudakov form factor is simply the probability of evolving from $Q_0$  to $Q$ without branching.
Eq.~\ref{foldcs} can be integrated to give an integral equation for $N(Q^2,\eta,k_\perp) $ in terms of the gluon TMD at the initial scale $Q_0$:
\begin{eqnarray}
N(Q^2,\eta,k_\perp) =N(Q_0^2,\eta,k_\perp) {\Delta_s(Q^2)} + \int ^{{Q}^2} _{Q^2_0} \frac{dt}{t} \frac{ \Delta_s(Q^2)}{\Delta_s(t)} \frac{\bar \alpha_s (t)}{2 \pi}
\int_{\Lambda_{\rm cut}}^{Q} \frac{ d^2 l_\perp }{l_\perp^2}N(t,\eta, k_\perp+l_\perp).
 \end{eqnarray}
 With the derived folded CS and renormalization group equation, we are ready to introduce the Monte Carlo implementation of the $k_t$ resummation formulated in the framework of the CGC effective theory.

\subsection{Forward evolution}

To have a consistency check, we first present the formulation of  the forward evolution scheme. The combined CS and renormalization group equation can be solved using the forward evolution approach. We lay out the main procedures in the following.

For a given virtuality scale $Q_i$, either after several steps of evolution or at the initial condition, we first generate the value of a higher virtuality scale $Q_{i+1}$, where the next branching occurs.
Following the conventional method, this can be achieved by solving the following equation,
 \begin{eqnarray}
\mathcal R=\exp \left[ - \int_{ Q_i^2}^{ Q_{i+1}^2}\frac{dt}{t} \bar \alpha_s(t) \left (\frac{1}{2}  \ln \frac{t}{\Lambda_{\rm cut}^2}-\beta_0\right ) \right ].
 \end{eqnarray}
where the argument of the running coupling $\alpha_s$ is simply chosen to be the virtual mass squared. 

Once $Q_{i+1}$ is generated, the transverse momentum of the radiated gluon, $l_{\perp,i+1}$, can be determined according to the following equation
\begin{eqnarray}
\mathcal R = \frac{1}{\cal C} \int_{\Lambda_{\rm cut}}^{l_{\perp,i+1}} \frac{ d^2 {l^\prime}_\perp }{{l^\prime}_\perp^2},
\end{eqnarray}
where the normalization factor reads ${\cal C} = \int_{\Lambda_{\rm cut}}^{Q_{i+1}} \frac{ d^2 {l^\prime}_\perp }{{l^\prime}_\perp^2}$. The four momenta of the radiated gluon and the $t$-channel gluon can be determined from the momentum conservation and the on-shell condition.
We will discuss the reconstruction of kinematics in more details in the next subsection.

 The generated cascade needs to be re-weighted. This is because that the unitary is no longer preserved beyond the leading double logarithm approximation. We have included the leading single logarithm contribution in the algorithm employed here,  which leads to the increase of   gluon number density after each splitting. The weighting factor is given by,
\begin{equation}
{\cal W} _{\rm CS} ( Q^2_{i+1}, Q^2_{i}) =  \frac{  \int ^{Q_{i+1}^2}_{Q_{i}^2} \frac{dt}{t}\alpha_s(t) \ln \frac{t}{\Lambda_{\rm cut}^2} } {\int ^{Q_{i+1}^2}_{Q_{i}^2}  \frac{dt}{t}\alpha_s(t)\left[ \ln \frac{t}{\Lambda_{\rm cut}^2} - 2\beta_0 \right] }.
\end{equation}
If the single logarithm contribution associated with the $\beta_0$ term in the denominator is neglected, the weighting factor reduces to 1.  With these re-weighted parton cascades, one can reconstruct the $t$-channel gluon $k_\perp$ distribution at different scales and compare with the analytical and numerical solutions of Eq.~\ref{comb}.

It is straightforward to numerically solve  Eq.~\ref{comb}, while the analytical solution of Eq.~\ref{comb} can also be easily obtained in the impact parameter space. After Fourier transforming back to the momentum space, the evolved gloun TMD distribution reads,
\begin{equation}
N(Q^2,\eta, k_\perp)=\int \frac{d^2 b_\perp}{(2\pi)^2} e^{i k_\perp \cdot b_\perp} e^{-S(\mu_b^2, Q^2)} \int d^2 l_\perp e^{-i l_\perp \cdot b_\perp } N(\eta, l_\perp) ,
\end{equation}
where $N(\eta, l_\perp)$ is  the gluon distribution evolved with the GLR equation, or the initial condition computed in the MV model. The Sudakov factor at one loop level in the impact parameter ($b_\perp$) space consists of a perturbative part and a non-perturbative part. It is given by
\begin{eqnarray}
S(\mu_b^2,Q^2)=  S_{pert}(\mu_{b*}^2,Q^2) +S_{NP} (b_\perp^2, Q^2).
\end{eqnarray}
The perturbative Sudakov factor reads
\begin{eqnarray}
 S_{pert}(\mu_{b*}^2,Q^2) =
\frac{N_c}{2\pi} \int^{Q^2}_{\mu_{b*}^2} \frac{d\mu^2}{\mu^2} \alpha_s(\mu) \left[ \ln
\frac{Q^2}{\mu^2} - 2 \beta_0 \right ],
\end{eqnarray}
where $\mu_{b*}^2$ is defined as $\mu_{b*}^2=4e^{-2\gamma_E}/b_{\perp*}^2$, with $b_{\perp *}=\frac{b_{\perp}}{\sqrt{1+b_{\perp}^2/b^2_{\max}}} $ and $b_{\max}=1.5\text{ GeV}^{-1}$. To compare with the Monte Carlo result on the same footing, we simply neglect the non-perturbative Sudakov factor $S_{NP}$ in the numerical calculation. The behaviour at large $b_\perp$ is regulated by $N(\eta, b_\perp)$ which is the Fourier transform of $N(\eta,l_\perp)$. In this work, we use the one-loop running coupling  which reads
\begin{equation}
    \alpha_s (\mu^2) = \frac{1}{\beta_{ 0}\frac{N_c}{\pi} \ln (\mu^2/\Lambda^2_{\rm QCD} )},
\end{equation}
with  $\Lambda^2_{\rm QCD}  = 0.0578$ GeV$^2$.

\begin{figure}
\centering
\includegraphics[width=0.4\textwidth]{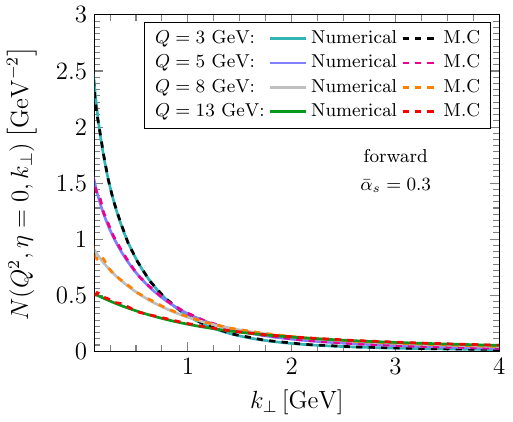}
\includegraphics[width=0.4\textwidth]{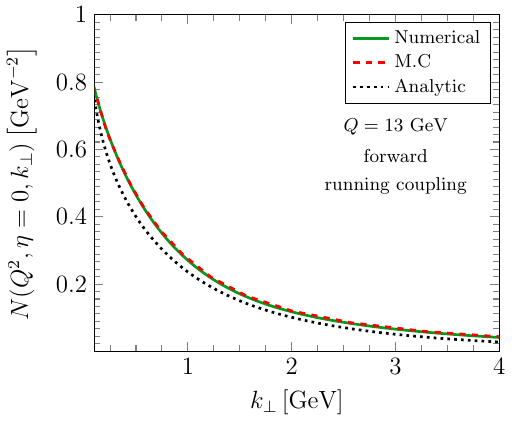}
    \caption{Comparison of the gluon $k_\perp$ distributions obtained from the forward evolution approach with the numerical solutions of the combined CS-renormalization group equation at different  scales(Color online). The left panel: the fixed coupling case. The right panel:  the running coupling case(Color online). }
    \label{for_cs}
\end{figure}

We present the $t$-channel gluon $k_\perp$ distribution constructed from the generated parton cascade and compare it with the numerical solution of the CS-renormalization group equation for the fixed coupling case in the left panel of Fig.~\ref{for_cs}. 
In our estimation,  the MV model is employed to provide with the gluon distribution at the initial scale  $Q_0=$3 GeV. In the formulation of TMD evolution, all soft-radiated gluons carry exactly zero longitudinal momentum fraction.  In contrast, all radiated soft gluons carry finite longitudinal momentum fraction in the parton branching algorithm. This presents an important advantage of the Monte Carlo method comparing with the conventional analytical approach. Keeping longitudinal momentum conservation exactly in parton splitting process is often crucial to correctly account for phenomenology near the threshold region~\cite{Liu:2022xsc}.   However, to make the comparisons in a consistent way, we didn't change the longitudinal momentum fraction of the $t$-channel gluon after each branching in our algorithm.  In the right panel of Fig.~\ref{for_cs}, we compare the Monte Carlo simulation result with both the numerical solution of the CS-renormalization equation and the analytical solution for the running coupling case at the scale $Q=13$ GeV. 
It is clear to see from the right panel of Fig.~\ref{for_cs} that our algorithm yields the same $k_\perp$ distribution as the numerical result. On the other hand, it differs from the analytical approach result. Such discrepancy is expected because the non-perturbative part of the CS kernel is treated differently in the analytical approach.  In addition, the argument of the running coupling used in the parton branching algorithm and the numerical solution is the hard scale $Q$, whereas the scale of running coupling is $\mu_b$ in the analytical approach. Since the analytical result can describe the relevant phenomenology very well, one should use it as guidance to model the non-perturbative part of the Sudakov factor which will be introduced in the Monte Carlo algorithm in the future work. Alternatively, one could also use a relatively large infrared cutoff value $\Lambda_{\rm cut}$ to mimic the effect of the non-perturbative Sudakov factor. We leave this for a future study.

\subsection{Backward evolution}

In this subsection, we will outline the essential steps of  Monte Carlo implementation for the backward evolution based on the folded CS-renormalization group evolution equation. Unlike the forward evolution which can be considered as a way of solving the evolution equation,  the evolved parton distributions have to be pre-generated and are used to guide the backward evolution.  In the most parton branching algorithm, the $k_t$ resummation is achieved by using the modified Sudakov factor incorporating the collinear Parton Distributions Functions (PDFs).  However, in the saturation regime, the $k_t$ resummation has to be formulated in terms of the unintegrated gluon distribution. The main procedures are summarized as follows.

The modified Sudakov factor in the backward evolution approach is different from that in the forward evolution approach. It reads
\begin{equation}
\Pi_{s}(Q_{i+1},Q_i; k_{\perp,i+1})=\frac{{\Delta_{s}( Q_{i+1}^2)} N(Q^2_{i}, \eta, k_{\perp,i+1}) }{{\Delta_{s}( Q_{i}^2)} N(Q^2_{i+1}, \eta, k_{\perp,i+1})}.
\label{backsdu1}
\end{equation}
An alternative way to compute the modified Sudakov factor is given by
\begin{eqnarray}
\Pi_{s}(Q_{i+1},Q_i; k_{\perp,i+1})
 &=& \exp\left [ -
 \int_{Q_{i}^2}^{Q_{i+1}^2} \frac{d t }{t} \frac{\bar \alpha_s(t)}{2\pi} \int^{ \sqrt t}_{\Lambda_{\rm cut}}  \frac{ d^2 l_{\perp} }{l_{\perp}^2 }  \frac{N(t,\eta, k_{\perp,i+1}+l_{\perp})}{N(t,\eta,  k_{\perp,i+1})} \right].  \label{backsdu2}
 \end{eqnarray}  
It describes the probability for gluon evolving backward from $Q_{i+1}$ to $Q_{i}$ without branching. The transverse momentum dependent gluon  distribution appearing in Eq.~\ref{backsdu1} and Eq.~\ref{backsdu2} has to be pre-generated by numerically solving the combined CS-renormalization group equation.

The backward evolution starts from the $t$-channel gluon with the highest virtuality $Q_{i}$. The hard scale of the partonic scattering process is denoted as $Q_{i+1}$.  We first have to sample $k_{\perp, i+1}$ according to the following distribution 
\begin{equation}
\mathcal R = \frac{1}{\cal C} \int ^{k_{\perp,i+1} }_{\Lambda_{\rm cut}}d^2k^\prime_\perp N(Q_{i+1}^2,\eta,k_{\perp}^\prime),
\end{equation}
with ${\cal C} = \int ^{Q_{i+1}}_{\Lambda_{\rm cut}} d^2 k^\prime_ \perp  N(Q_{i+1}^2,\eta,k_{\perp}^\prime)$ being the normalization factor. The rapidity $\eta$ is fixed by external kinematics. The next quantity to be generated by the parton cascade algorithm is the value of   virtuality $Q_{i}$. 

Following the standard backward evolution strategy, $Q_{i}$ is obtained using the backward type Sudakov factor. We can sample a $Q_{i}$ by solving the following equation,
\begin{eqnarray}
{\cal R} = \Pi_{s}(Q_{i+1},Q_{i}; k_{\perp,i+1}). \label{sback}
\end{eqnarray}

As the virtual mass of $(i+1)$th $t$-channel gluon, $Q_i$ also serves as the hard probe scale at which the $i$th $t$-channel gluon's transverse momentum is measured. The transverse momentum of the radiated gluon $l_{\perp,i}$ is thus sampled solving the following equation
\begin{align}
& \mathcal R  = \frac{1}{\cal C} \int ^{l_{\perp,i}}_{\Lambda_{\rm cut}} \frac{d^2l'_\perp}{l'^2_\perp}N(Q_i^2,\eta,k_{\perp,i+1 }+l'_\perp),
\\
&{\cal C} =  \int ^{Q_{i}}_{\Lambda_{\rm cut}} \frac{d^2l'_\perp}{l'^2_\perp}N(Q_i^2,\eta,k_{\perp,i+1}+l'_\perp) .
\end{align}
The longitudinal momentum fraction of the radiated gluon is determined through the on-shell condition,
\begin{equation}
|Q_i^2|\approx \frac{z_i l_{\perp,i}^2}{1-z_i}+|k_{\perp,i+1}^2|,
\end{equation}
which is valid in the strong ordering region $|Q_{i-1}^2 | \ll |Q_i^2| \ll |Q_{i+1}^2|$. The minus component of the emitted gluon can be fixed accordingly. The $i$th $t$-channel gluon's transverse momentum is trivially obtained:  $k_{\perp,i}=k_{\perp,i+1}-l_{\perp,i}$.  The virtual mass $Q_{i-1}$ of the $i$th $t$-channel gluon is computed with Eq.~\ref{sback}.  However, $t$-channel gluons' four momenta can be determined only after the whole cascade is generated. The minus component of the $t$-channel gluon that is directly attached to nuclear target is set to be 0.  From this initial condition, the four momenta of $t$-channel gluons are retrospectively reconstructed by momentum conservation. 
 
As argued in the previous subsection, the generated event has to be re-weighted after each branching since the unitary is not preserved in the single leading logarithm accuracy level. In the backward evolution approach, the re-weighting function reads
 \begin{equation}
{\cal W} _{\rm CS, back} ( Q^2_{i+1}, Q^2_{i}) = \frac{\int ^{Q_{i+1}^2}_{Q_{i}^2}  \frac{dt}{t}\alpha_s(t) \left[ \ln \frac{t}{\Lambda_{\rm cut}^2}- 2\beta_0 \right] } { \int ^{Q_{i+1}^2}_{Q_{i}^2} \frac{dt}{t}\alpha_s(t) \ln \frac{t}{\Lambda_{\rm cut}^2} } \ .
\end{equation}


We repeat the procedure outlined above until $Q_{i}^2$ reach  a minimal  cut-off scale at which TMD evolution stops.  The TMD evolution is driven by the soft gluon radiations which carry the vanishing longitudinal momentum fraction $1-z_i \rightarrow 0$. In the practical Monte Carlo implementation, the cut-off is chosen to be $|Q_{i}^2|>|l_{\perp,i}^2|+|k_{\perp,i+1}^2| $, or equivalently $z_i>0.5 $.  Meanwhile, $|Q_{i}^2|$ is also required to be larger than the satuartion scale $Q_s^2$.  If these two conditions can not be met simultaneously, we terminate  the TMD evolution, and start the backward small $x$ evolution.

\begin{figure}[htb]
\centering
\includegraphics[width=0.4\textwidth]{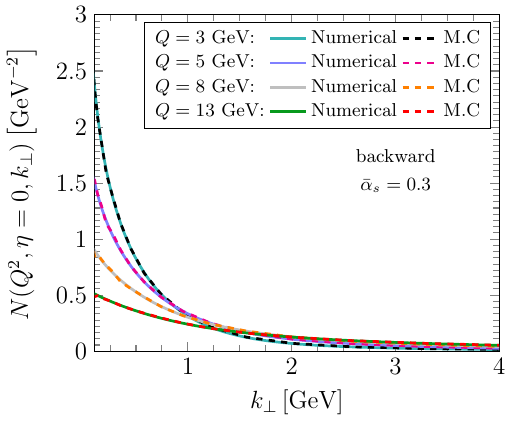}
\includegraphics[width=0.4\textwidth]{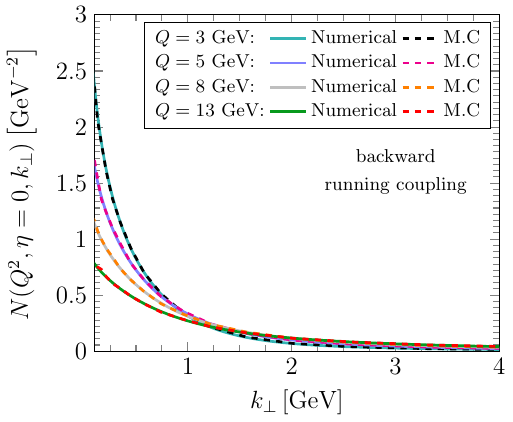}
    \caption{Comparison of the gluon $k_\perp$ distributions obtained from the backward  approach with the numerical solutions of the CS-renormalzation group equation at different  scales (Color online).}
    \label{back_cs}
\end{figure}

We test the backward evolution  algorithm against the numerical method as shown in Fig.~\ref{back_cs}. The MV model result is applied at the initial scale $Q_0$=3 GeV. The gluon $k_\perp$ distribution at high scale $Q=13$ GeV is obtained by numerically solving the combined CS-renormalization group equation. The cascade  is generated starting from the scale $Q=13$ GeV and evolve down to the initial scale with the backward approach. The $t$-channel  gluon $k_\perp$ distribution reconstructed from the cascade is compared with the numerical results at different scales.
Gluon $k_\perp $ distributions are presented  in the left panel of Fig.~\ref{back_cs} for the fixed coupling case,  and in the right panel of Fig.~\ref{back_cs} for the running coupling case. It is evident that the $k_\perp$ distributions obtained from the Monte Carlo method is the same as the numerical results.
We conclude that the backward evolution algorithm pass this consistency check as expected.

\section{Conclusion}
In this work, we extended the small $x$ initial state parton branching algorithm developed in the previous paper to include the kinematic constraint effect. In the small $x$ limit, the kinematic constraint leads to stronger suppression of soft gluon emissions than that caused by the angular ordering along the chain. The coherent branching effect is thus effectively implemented in the parton branching algorithm once the kinematic constraint is imposed. This is a nontrivial extension in the sense that the weighting factor and the way of sampling radiated gluon's transverse momenta are drastically altered. The $t$-channel gluon $k_\perp$  distributions constructed from both the forward scheme and the backward scheme are shown to reproduce the numerical solutions of the kinematic constrained GLR equation.

We also formulated a parton branching algorithm that enables us to resum large $k_t$ logarithms at small $x$ logarithms following a two-step evolution picture. The cascade first develops by radiating soft gluons that carry vanishing longitudinal momentum fractions in the backward approach description.  At this first stage of the evolution, the parton branching is simulated with the Sudakov factor which we obtained from the folded CS equation and the renormalization group equation.  The transverse momentum-dependent gluon distribution instead of gluon PDF is used to guide the evolution path toward the most populated regions of $(Q^2, k_\perp)$.
When the virtual mass of the $t$-channel gluon is dominated by its transverse momentum or is of the order of saturation scale, the parton branching starts being generated according to the non-Sudakov form factor derived from the small $x$ evolution equation. The joint $k_t$ and small $x$ resummation thus has been achieved in the Monte Carlo simulation by implementing such two-step evolution. Our study represents an important step towards practical applications of the parton shower generator in simulating scattering processes that involve multiple well-separated hard scales, such as di-jet production in eA collisions at EIC.   The next step is to construct a  full hadron-level Monte Carlo generator with the hadronization being performed using multi-purpose generators such as PYTHIA~\cite{Sjostrand:2006za}.  We also plan to integrate the algorithm into eHIJING framework~\cite{Ke:2023xeo} aiming at the simulation of events in eA collisions for the whole $x$ range accessible at EIC in the future.

\  

{\it Acknowledgments:}
We thank Hai-tao Li and Shan-shan Cao for helpful discussions.
This work has been supported by the National Natural Science Foundation of China under Grant No. 1217511. Y.S. is supported by the China Postdoctoral Science Foundation under Grant No. 2022M720082. S.Y.W. is also supported by the Taishan fellowship of Shandong Province for junior scientists.  




\bibliographystyle{apsrev4-1}
\bibliography{ref.bib}

\end{document}